\documentclass{openjournal}

\usepackage{xcolor}
\usepackage{textgreek}
\usepackage[utf8]{inputenc}
\usepackage[english]{babel}
\usepackage{setspace}
\usepackage[colorlinks=true, allcolors=blue]{hyperref}
\hypersetup{
    unicode,
    colorlinks=true,
    linkcolor=linkcolor,
    citecolor=linkcolor,
    filecolor=linkcolor,
    urlcolor=linkcolor,
}
\usepackage{color,colortbl}
\definecolor{linkcolor}{rgb}{0.0,0.3,0.5}
\usepackage{tensind}
\tensordelimiter{?}
\DeclareGraphicsExtensions{.bmp,.png,.jpg,.pdf}
\usepackage{verbatim}
\usepackage[normalem]{ulem}
\usepackage{orcidlink}
\usepackage{soul}
\usepackage{acro}
\usepackage{xspace}
\usepackage{amsmath,amsfonts,amssymb}
\usepackage{graphicx}
\usepackage{relsize}
\usepackage{acro}
\usepackage{listings}
\usepackage{xcolor, fontawesome}
\usepackage{multirow}

\usepackage{natbib}
\usepackage{longtable}

\acsetup{
  single    = false,
  list/sort  = true,
  cite/group = true,
  cite/group/cmd = \citealt,
  cite/group/pre = {; \xspace},
  patch/longtable=false,
}

\DeclareAcronym{psf}{
  short = PSF,
  long = point spread function,
}

\DeclareAcronym{fim}{
  short = FIM,
  long = Fisher Information Matrix,
  cite = {fisher1925}
}

\DeclareAcronym{hmc}{
  short = HMC,
  long = Hamiltonian Monte Carlo,
  cite = {Betancourt2017}
}

\DeclareAcronym{crlb}{
  short = CRLB,
  long = Cram\'{e}r-Rao Lower Bound,
  cite = {rao1945,cramer1947}
}

\newcommand\jax{\textsc{Jax}\xspace}
\newcommand\dlux{$\partial$Lux\xspace}

\newcommand\optax{\texttt{optax}\xspace}
\newcommand\numpyro{\texttt{numpyro}\xspace}

\newcommand\loglike{$\mathcal{L}$\xspace}

\graphicspath{ {./figs/} }

\begin{document}
\title{Differentiable Optics with $\partial$Lux II: Optical Design Maximizing Fisher Information}

\author{Louis Desdoigts}
\email{louis.desdoigts@sydney.edu.au}
\affiliation{Leiden Observatory, Niels Bohrweg 2, Leiden 2300RA, The Netherlands}
\affiliation{Sydney Institute for Astronomy, School of Physics, University of Sydney, NSW~2006, Australia}

\author{Benjamin J. S. Pope}
\affiliation{School of Mathematical \& Physical Sciences, Macquarie University, 12 Wally's Walk, Macquarie Park, NSW 2113, Australia}
\affiliation{School of Mathematics and Physics, University of Queensland, St Lucia, QLD~4072, Australia}

\author{Michael Gully-Santiago}
\affiliation{University of Texas at Austin, Department of Astronomy, 2515 Speedway, Stop C1400, Austin, Texas 78712-1205, USA}

\author{Peter G. Tuthill}
\affiliation{Sydney Institute for Astronomy, School of Physics, University of Sydney, NSW~2006, Australia}

\begin{abstract}

The design of astronomical hardware operating at the diffraction limit requires optimisation of physical optical simulations of the instrument with respect to desired figures of merit, such as photometric or astrometric precision.
System design entails many parameters some of which may entangle the fidelity of science observables with strongly nonlinear dependencies upon instrument properties.
Here we present a differentiable optical simulation framework \href{https://github.com/LouisDesdoigts/dLux}{\dlux}, a software library designed to construct optical models that are integrated with automatic differentiation. This approach enables the direct evaluation of gradients and higher-order derivatives through the forward model, facilitating statistically principled design and optimisation of instrument configurations.
The methodology leverages numerically stable second- and higher-order derivatives to directly compute Fisher information and covariance forecasts, enabling efficient Bayesian experimental design targeted toward optimising abstracted figures of merit, such as the precision of parameters recovered through complex sets of operations.
The method is validated against analytical results and applied to optimise the astrometric precision achievable with a parametrised telescope model and a diffractive pupil design relevant to exoplanet detection missions.
To support reproducibility and facilitate methodological extension, we provide example implementations via open-source code on the Github sharing platform\footnote{\href{https://github.com/LouisDesdoigts/FIM_tutorial/blob/main/tutorial.ipynb}{github.com/LouisDesdoigts/FIM\_tutorial/blob/main/tutorial.ipynb}}.

\end{abstract}


\maketitle

\begin{spacing}{1.15}   %

\section{Introduction}
\label{sec:intro}

Advances in contemporary observational astronomy are driven by progress in instrumentation and hardware, enabling measurements with improved sensitivity and lower noise. This is particularly true for exoplanetary science where signals can be many orders of magnitude smaller than the noise. When designing instrumentation for this goal, it is essential to accurately model the pattern, or \ac{psf}, governing the spread of starlight on the detector to enable the engineering of system hardware to optimise science deliverables.
For example, apodizing phase plate coronagraphs~\citep{Guyon2006} suppress light from on-axis sources by modifying the wavefront in multiple conjugate planes, creating spatially-varying PSFs. This transformation allows faint nearby companions to be detected against the glare of their bright host star.
This can be achieved with an apodizing phase plate \citep{Codona2006}, a phase mask placed in a telescope pupil plane that reshapes the \ac{psf}, the design of which poses nonlinear optimisation challenges \citep{por2017}. Astrometric exoplanet detection \citep{Sozzetti2005} similarly requires careful consideration of instrumental design such that \acp{psf} are stable and can be used to characterise the time-varying instrumental imperfections, achieving measurement precisions required for tiny (micro-arcsecond) signals. This can be achieved with widefield space-based surveys such as \textit{Gaia}, or with smaller telescopes that employ a diffractive pupil to engineer \acp{psf} with favourable properties \citep{guyon2012,tuthill2018}. %
In either case and throughout observational astronomy we find escalating demands for stability and precision in data-driven calibration of diffraction effects in the \ac{psf}.

Advances in software and algorithms are therefore essential for both data analysis and to facilitate hardware design. A range of open-source physical optics simulation codes \citep{Perrin2012,prysm,hcipy} have been developed to model imaging systems end to end, providing a framework to optimise design and/or data analysis schemes by way of grid-based or Markov Chain Monte Carlo~\citep{{Metropolis1953}} methods. For models with many parameters (the norm for realistic situations with, for example, many modes of phase aberration, multiple sources in the field of view and/or pixel-level imperfections in the detector) then reliable sampling or optimisation can become computationally intractable \citep{Huijser2022} unless it is also possible to evaluate not just the value but also the \textit{gradient} of the objective function.

Algorithms to implement automatic differentiation or `autodiff' \citep{Margossian2018} comprise the foundational technology used in artificial intelligence and machine learning \citep{lecun15}, and are now rapidly becoming a critical enabling technology in software for the physical sciences. Autodiff makes it possible to evaluate the partial derivatives of a computer program's floating-point outputs with respect to floating-point arguments.
Autodiff computes exact derivatives using successive applications of the chain rule, with computational cost scaling similarly to that of the original model evaluation, rather than increasing with the number of parameters.
Considerable industry and academic effort has gone into developing performant and user-friendly numerical software libraries with autodiff capability: \textsc{PyTorch} \citep{pytorch}, \textsc{TensorFlow} \citep{tensorflow2015}, Julia \citep{julia}, or our preferred library used in this work, \jax \citep{jax}. Importantly, autodiff removes the need to manually derive or symbolically express gradients or Hessians --- they emerge directly from the differentiable structure of the implemented forward model. This is a central departure from traditional inference workflows, and it underpins our ability to evaluate and optimise realistic optical systems without requiring closed-form expressions for likelihood derivatives.

Differentiable forward models in physics permit optimisation and inference with very many parameters, and the inclusion of flexible nonparametric models \citep{lavin2022simulation} jointly with deterministic physics. %
Autodiff has been used recently in optical and imaging science for phase retrieval and \ac{psf} modelling \citep{jurling_fienup,phase_ret_and_design,liaudat2021,Liaudat2023}.
In previous work~\citep{Desdoigts2023} we introduced \dlux, a differentiable physical optics framework designed to address a variety of high dimensional optical calibration tasks. Using the accelerated numerical computation methods and gradient calculations enabled by its \textsc{Jax} back-end, efficient solutions to complex optical problems such as end-to-end phase retrieval and pixel-level calibration from simulated imaging data were demonstrated.

In this manuscript we build on this previous work and focus on scenarios where the posterior distribution over model parameters is well described by a multivariate normal near its peak, known as the Laplace approximation \citep{Kass1991,mackay2002}. We show how numerically stable higher-order gradients permit the calculation of covariance and Fisher matrices \citep{Bhandari2021,Coulton2023} --- a key object in statistical design that quantifies how precisely model parameters can be inferred from observations --- under the Laplace approximation. By constructing differentiable optical models we can differentiate a likelihood function with respect to any astrophysical or instrumental model to compute parameter covariance matrices. This means that neither the likelihood function nor its derivatives need to be written in closed form. Instead, they are evaluated programmatically via autodiff directly through the forward model, a key methodological shift enabled by this approach. This enables {\it Fisher forecasting}: a statistical technique used to estimate the precision of parameter estimates from a future experiment, and enables computation of the covariance matrix. The covariance matrix obtained through the Fisher matrix gives us the \ac{crlb} --- the information-theoretical limit on parameter constraints achievable with an experiment. This statistically principled approach is common for the planning of new instruments and surveys; for example in cosmology the forecasting of how well observations of the cosmic microwave background \citep{Liu2016} or spectroscopic surveys \citep{d'AssigniesD2023} will constrain bulk cosmological parameters, under varying choice of settings in a preliminary instrument design, and marginalised over nuisance parameters.

The ability to calculate these Fisher matrices using autodiff then enables {\it Bayesian experimental design} \citep{fedorov1972,chaloner95,ryan2016} via gradient descent. This might involve optimising the uncertainty of a particular parameter, or a norm such as the determinant or trace of the \ac{crlb}. In Section~\ref{sec:fisher} of this Paper we briefly describe the underlying theory, and in Section~\ref{sec:theory} we validate calculations made with \dlux against analytic theory. By way of illustration with a concrete example, Section~\ref{sec:optimization} applies this framework to optimise the design of a phase mask for the astrometric mission \textsc{Toliman} \citep{tuthill2018}, a small telescope aiming to detect planets around $\alpha$~Cen AB by measuring micro-arcsecond perturbations in the relative separation of the binary.

\newpage
\section{Fisher Information and Bayesian Experimental Design}
\label{sec:fisher}

The key feature of autodiff exploited in this paper is the ability to efficiently calculate a multivariate normal approximation to a probability distribution. Consider an imaging system and sources, parametrised by %
$\theta \equiv (\theta^1, ... \theta^N)$, that generate data $d$. In solving an inverse problem, we want to infer $\theta$ given $d$ and any prior information. We can use Bayes' rule to update our prior knowledge of $\theta$ conditioned on our prior knowledge $I$, given $d$

\begin{equation}
    \overbrace{p({\theta}|d,I)}^\text{posterior} = \frac{{\overbrace{p(d|{\theta},I)}^{\text{likelihood}} \cdot \overbrace{p({\theta}|I)}^{\text{prior}}}}{{\underbrace{p(d|I)}_{\text{evidence}}}}.
\end{equation}

\noindent It is convenient computationally to express this in logarithmic units

\begin{equation}
    \log p({\theta}|d,I) = \overbrace{\mathcal{L}(d|{\theta},I)}^\text{log-likelihood} +
    \overbrace{\Pi({\theta}|I)}^\text{log-prior} - \overbrace{\log Z}^\text{log-evidence}.
\end{equation}

At the maximum likelihood, the \textit{score} --- the gradient of the log-likelihood with respect to the parameters --- vanishes, and %
$\partial_i\mathcal{L}({\theta}_0) = 0 \ \forall _i$ where $\partial_i \equiv \partial/\partial\theta^i$ is the maximum likelihood estimate of these parameters. We then motivate consideration of a Taylor expansion of \loglike about this point, dropping the prior $\Pi$\footnote{We can do so without loss of generality; to include the prior, we can replace \loglike throughout this calculation with $\mathcal{L} + \Pi$ instead.}, to estimate the distribution near the maximum. Going up to second-order is equivalent to approximating \loglike as a multivariate normal --- ie Laplace's Method

\begin{align}
    \mathcal{L}({\theta^i}) \approx\ \mathcal{L}({\theta}^i_0) + (\theta - \theta_0)^i\underbrace{\frac{\partial\mathcal{L}}{\partial\theta^i}}_{=0}\Bigg|_{{\theta=\theta}_0} \nonumber + (\theta - \theta_0)^i(\theta - \theta_0)^j\underbrace{\frac{\partial^2\mathcal{L}}{\partial\theta^i\partial\theta^j}}_{\equiv F_{ij}}\Bigg|_{{\theta=\theta}_0} + \underbrace{\cdots}_{\text{higher-order terms}}
\end{align}

\noindent and we identify the negative Hessian of \loglike as the \ac{fim}

\begin{equation}
    \mathbf{F}(\boldsymbol\theta) \equiv F_{ij}(\boldsymbol{\theta}) = -\frac{\partial^2\mathcal{L}}{\partial\theta^i\partial\theta^j}.
\end{equation}

\noindent This is formally defined as the variance of the score, and for appropriately regular functions is given by this Hessian. This second-order approximation treats the log-likelihood as a quadratic function near the maximum, and is often used in practice when full posterior sampling is computationally prohibitive. The parameter covariance matrix $\mathbf{C}$ is calculated as the inverse of the \ac{fim}

\begin{equation}
    \mathbf{C} = \mathbf{F} ^ {-1}
\end{equation}

\noindent that fully describes the best-fit multivariate normal; an estimator of the behaviour of a likelihood distribution around its peak. In classical applications, this would require symbolic or manual calculation of second derivatives of the log-likelihood. However, in our framework, these quantities are obtained directly via automatic differentiation of the implemented model, without the need for algebraic derivation.

The \ac{fim} has some convenient properties \citep{Coe2009}, firstly the Fisher matrices for two independent experiments add ($\mathbf{F}_{1,2} = \mathbf{F}_1 + \mathbf{F}_2$) allowing for the \ac{fim} to be calculated independently for any observations that add linearly, such as dithered images.
Secondly, since parameter marginalisation happens through the matrix inversion of $\mathbf{F}$, row and column $i$ can be deleted from $\mathbf{F}$ in order to remove its contribution to the resulting covariance matrix.
The covariance matrix of a model with parameters $\boldsymbol{\theta}$ is the \ac{crlb}: the lower bound on the variance of an unbiased frequentist estimator of $\boldsymbol{\theta}$.
It is not possible to recover parameters better than the \ac{crlb}, so it is useful for forecasting the sensitivity of an experiment to the parameters of a model under consideration, irrespective of how data analysis will be carried out.

In the remainder of this paper, we will assume that the multivariate normal approximation to our likelihood holds, and that the Fisher and covariance matrices calculated from this Hessian are accurate to within a reasonable tolerance. We also assume that our measurements are photon-noise dominated and therefore  a Poissonian likelihood  is used exclusively throughout the remainder of the work. Furthermore, all calculations of, and references to, the covariance matrix are of the parameter covariance matrix, calculated against simulated data without noise realisations, and therefore completely describe the \ac{crlb} under the Laplace approximation (which is assumed valid throughout).

\section{Comparison with Theory}
\label{sec:theory}

In especially simple cases and under a number of simplifying assumptions, the Fisher forecast of parameters of interest can be obtained by analytic theory. In this Section we compare the compare the estimated \ac{crlb} uncertainties via three different methods: an analytic derivation, through the Fisher matrix calculated via a differentiable model (the methodology explored in this paper), and direct sampling of the posterior via an \ac{hmc} algorithm. The posterior samples generated via the \ac{hmc} serve as a numerical benchmark we can compare the Fisher and analytic methods against, given that the \ac{hmc} directly samples the full posterior.

\begin{figure*}%
    \centering
    \includegraphics[width=1.0\textwidth]{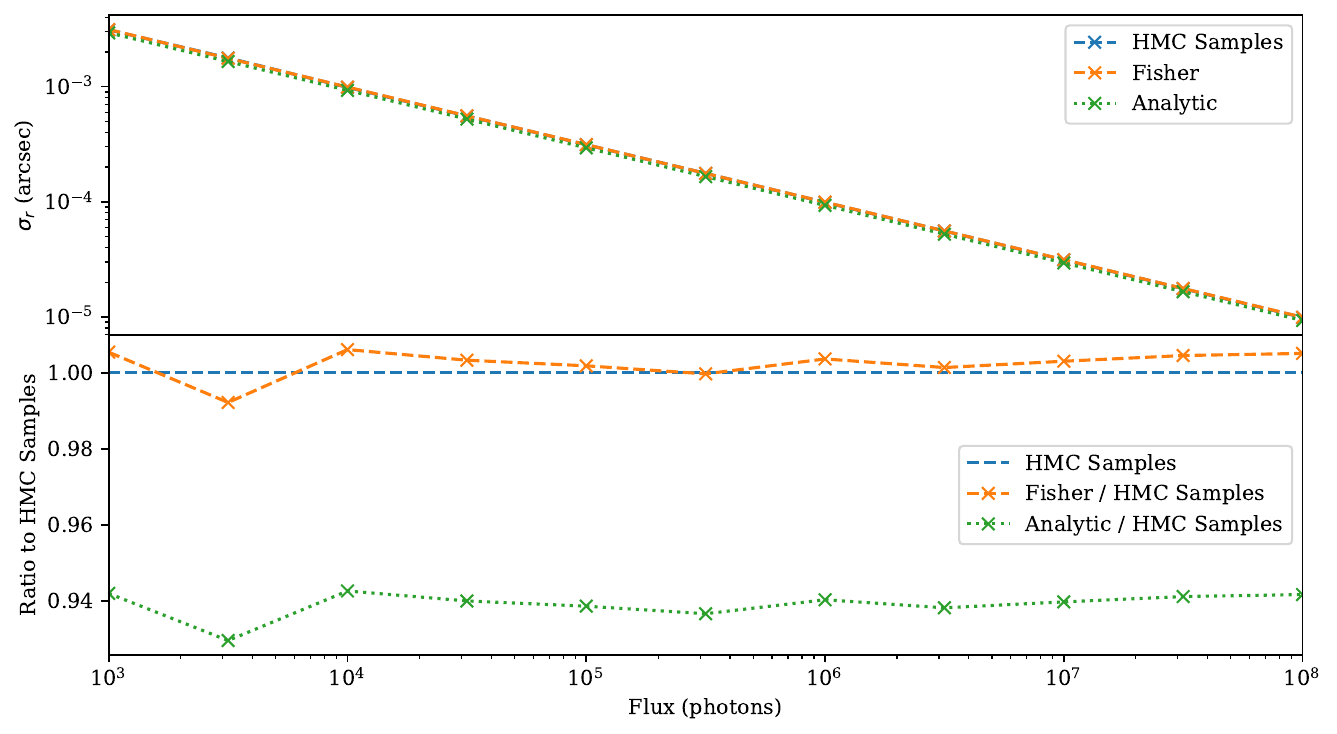}
    \caption[Analytic versus Autodiff Comparison]{Comparison of the posterior positional uncertainty, $\sigma_r$, via three estimators: \ac{hmc} posterior samples (used as a numerical benchmark), a Fisher-information estimate obtained via autodiff, and a closed form analytic derivation. Top: absolute estimated $\sigma_r$ versus photon count for the three estimators, all showing strong agreement. Bottom: ratio of $\sigma_r$ to the \ac{hmc} benchmark (unity = perfect agreement), revealing a systematic difference between the analytic method and the \ac{hmc} benchmark of $\sim6\%$, indicative of an consistent overestimate of system performance. The Fisher estimate is in strong agreement with the \ac{hmc} benchmark at all values.}
    \label{fig:theory_comparison}
\end{figure*}

We start with an illustrative toy problem: obtaining a position measurement from the image plane of a simple telescope. Given a circular pupil support and a monochromatic point source, the \ac{psf} is the well known Airy disk. The analytic \ac{crlb} (i.e. best achievable precision, independent of the analysis method used) on the localisation of a point source through this system as a function of photon count is given by the expression

\begin{equation}
\label{eq:r_std}
    \sigma_{r} = \frac{1}{\pi} \sqrt{\frac{2}{N_\text{phot}}}  \frac{\lambda}{D},
\end{equation}

\noindent where $\sigma_r$ is the radial uncertainty in the positional measurement in radians, $N_\text{phot}$ is the number of photons, $\lambda$ is the wavelength of light and $D$ is the diameter of the aperture \citep{guyon2012}. Figure~\ref{fig:theory_comparison} now compares $\sigma_r$ as calculated from this analytic equation and the $\sigma_r$ values calculated using the Fisher methods described in Section~\ref{sec:fisher}, both registered against results of the posterior samples generated via the \ac{hmc} algorithm, used as the benchmark. We assumed that the signal was photon noise limited, and therefore chose a per-pixel Poisson distribution as the likelihood function for both the Fisher and \ac{hmc} methods, which is the same as that used to derive Equation~\ref{eq:r_std}.   %

The top panel of Figure~\ref{fig:theory_comparison} shows consistency between the three methods across the full range of the fluxes, however the bottom panel that compares the Fisher and analytic method to the \ac{hmc} numerical benchmark shows a consistent overestimate of performance with the analytic approach. This highlights the limitations of the idealised assumptions typically required to find closed-form solutions to these equations. In this case, the finite pixel and image plane size form two such complexities not addressed by analytic derivation. Furthermore, any real-world imaging system is likely to entail significantly more complexity than this toy model. Instruments operate over a non-zero passband and feature chromatic performance, surfaces are imperfect, and the clear aperture can be cluttered with blockages from the secondary mirror and spiders. All these, and many more non-ideal elements will directly influence the \acp{psf} and in general can only be modelled numerically. In these cases, differentiable forward models provide an effective way to accurately calculate the \ac{crlb} of a given observation without resort to the simplifying assumptions required for analytic expressions.

\section{Optical Design Case Study: the TOLIMAN diffractive pupil}
\label{sec:optimization}

These methods allow the evaluation of the covariance matrix at fixed parameter values and the direct optimisation of its components using higher-order derivatives. This enables gradient-based experimental design, targeting properties of the covariance matrix such as entropy or marginalised variances, using a differentiable forward model.
This opens up the possibility of \emph{Bayesian experimental design}, allowing for gradient descent directly on such properties and functions of the covariance matrix, or as we show in this example, on individual components of the covariance matrix marginalised over the rest of the model. In keeping with the central theme of this paper, accomplishing this requires only a forward model of the system in a differentiable framework. %

The illustrative optical problem chosen is the design of the diffractive pupil required for the \textsc{Toliman} mission \citep{tuthill2018}, a $\sim$100\,mm class aperture space telescope which aims to measure the micro-arcsecond scale angular perturbation induced on the host star by the orbital gravitational reflex motion from an (unseen) low-mass planet.
Such a formidable challenge requires the engineering of a \ac{psf} that maximally encodes both instrument metrology and scientific signals of interest simultaneously on the sensor.
The solution to these challenges incorporates a binary-valued diffractive pupil and a spectrometer, all integrated into the \ac{psf} of the telescope by engraving phase patterns onto the entrance pupil. Specifically crossed sinusoidal gratings are used to produce a spectrometer (not discussed further in the present manuscript), which are overwritten onto the primary diffractive pupil pattern which enables the optical and detector calibration while also allowing astrometric science measurement. Here we focus solely on this latter pattern which consists of regions of zero/$\pi$ phase which are separated by sharp transitional boundaries.  The requirement for the binary pattern arises as a primary design constraint of the \textsc{Toliman} diffractive pupil~\citep{tuthill2018}.
The optical signal encoding and associated recovery methods for system state metrology for the \textsc{Toliman} mission present complex requirements, providing an ideal robust test case for evaluating the practical capabilities and limitations of the Fisher-information-based design methods discussed in this paper. While full details of the mission-specific constraints are not essential here, this context highlights the degree of design complexity and abstracted figures of merit that can be effectively managed through differentiable optical modelling.

\begin{figure}[!h]
    \vspace{25pt}
    \centering
    \includegraphics[width=1.\linewidth]{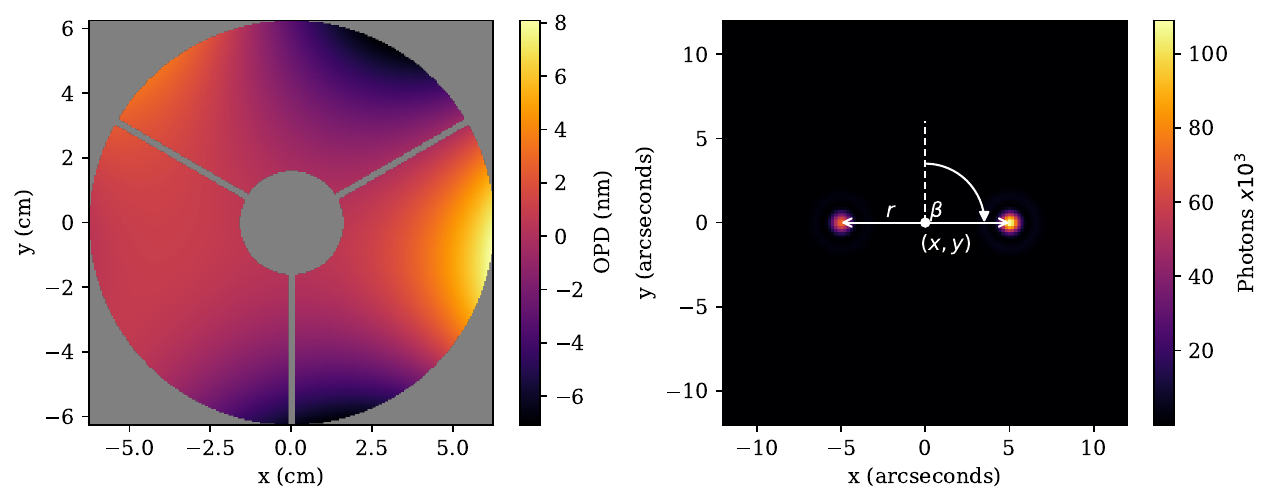}
    \caption{Geometric parameters and Toliman astrometric problem. Left: telescope aperture geometry with random phase errors generated on the aperture. Optical aberrations ${Z_i}$ are represented with Zernike polynomials in units of nm. Right: parametrisation of the binary star geometry; mean position $(x, y)$ (arcseconds), relative separation $r$ (arcseconds), position angle $\beta$ (degrees).\vspace{25pt}}%
    \label{fig:toliman_geometry}
\end{figure}

Here, we take the design of the \textsc{Toliman} diffractive pupil as a complex yet achievable demonstration of analysis by our autodiff frameworks, and further posit that arriving at any comparably performing outcome by way of analytic or heuristic approaches would present a formidable task.
We seek a binary-valued diffractive pupil that maximally constrains the binary separation $r$, marginalising over the remaining astrophysical parameters: the mean position $(x, y)$ in arcseconds on the sensor, the position angle $\beta$ in degrees, the total flux $F$ in photons (parametrised as $\log_{10}(F)$ in the Fisher and covariance calculations), a dimensionless contrast $\phi$ as the ratio of the binary component fluxes, the mean wavelength $\lambda$ in nm, as well as the remaining optical parameters --- the pixel scale $\gamma$ in arcsec/pixel and optical aberrations ${Z_i}$ modelled as a sum of normalised Zernike polynomials with coefficients in nm. Figure~\ref{fig:toliman_geometry} shows the basic geometry of both the telescope aperture and the astrophysical system, without any diffractive pupil in the system.

TOLIMAN's requirement for binary phases $\in \{0,\pi\}$ presents a challenge for gradient-based optimisation methods as the distribution is inherently discontinuous. To map differentiably between a continuous set of basis vectors and binary valued mask we employ the CLIMB algorithm, first introduced for the Toliman mask design problem in \citet{phase_ret_and_design}. By using a set of 3$\times$ oversampled basis vectors, a binary mask can be constructed with a single pixel boundary between the two regions with continuous values that enable gradients to propagate smoothly. In brief, any 3$\times$3 oversampled pixel with all values above or below zero can be assigned to one or zero respectively. Any remaining pixels form the boundary region, where a plane can be fit via least-squares to approximate the fraction of the pixel above zero, yielding a single pixel boundary of continuous values between zero and one. More details and full psuedo-code can be found in \citet{phase_ret_and_design}.

\begin{figure}

    \centering
    \includegraphics[width=1.\linewidth]{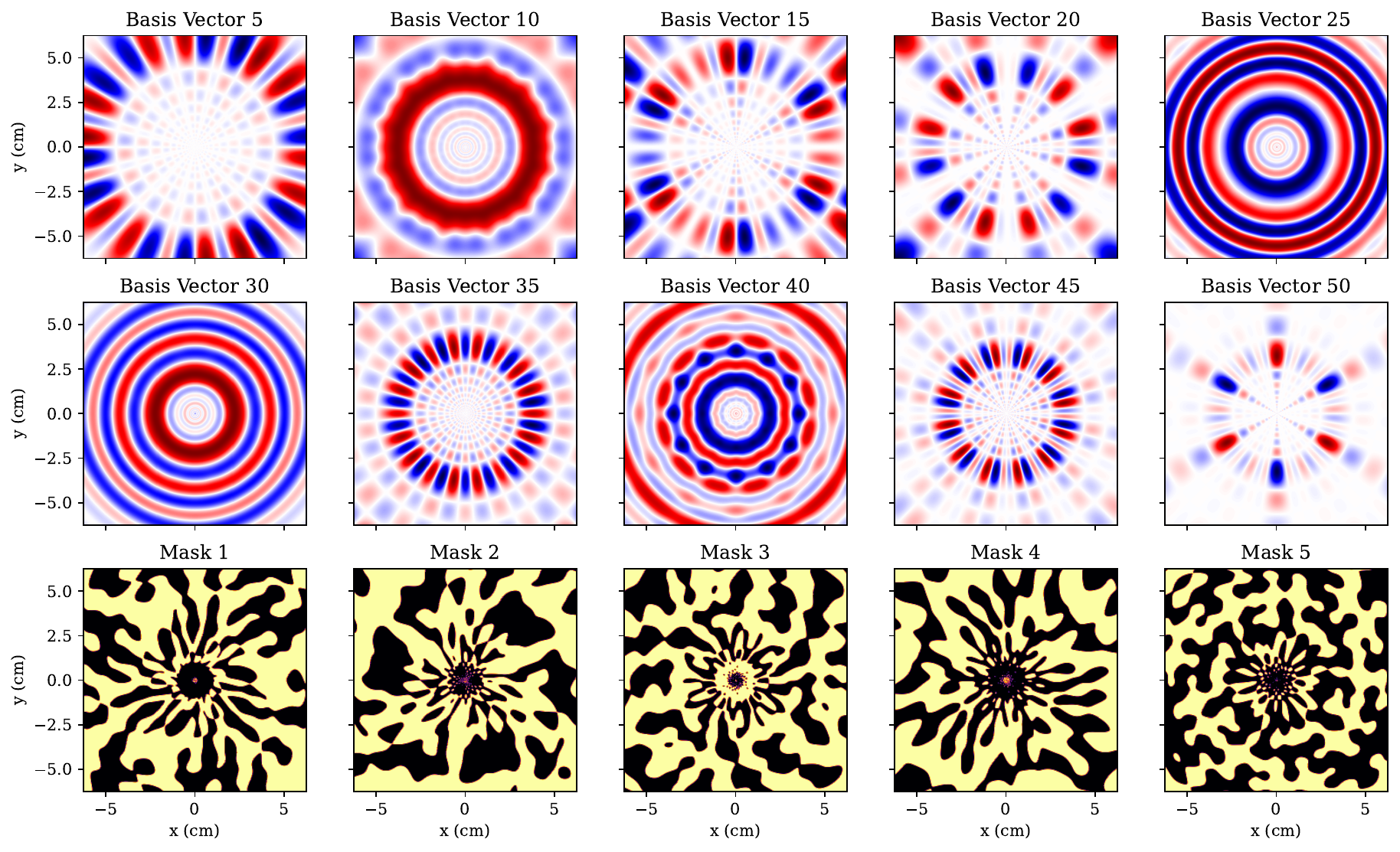}
    \caption{Top two panels: selection of the the ortho-normalised continues basis vectors built from log-harmonics and radial sine and cosine functions. Colour scales are arbitrary. Bottom panel: Binary masks generated by random coefficients. The dark and light regions correspond to zero and $\pi$ respectively.\vspace{25pt}}
    \label{fig:basis_vectors}
\end{figure}

The basis vectors used can, in principle, be chosen freely. However, it is advantageous to tailor their selection to the physical symmetries of the problem. In this work, we employ basis vectors with three-fold rotational symmetry, which naturally enhances sensitivity to even-mode Zernike aberrations that suffer sign degeneracies in conventional optical systems. By improving the sensitivity to these modes we enable the diffractive pupil to both constrain the astrophysical geometry and act as a wavefront sensor. Ultimately the choice of basis vectors is arbitrary as the differentiable engine propagates gradients from \ac{psf}s back to the basis of choice. In this work we build the basis vectors from a set of log harmonic radial and radial sine and cosine functions orthogonalised using the Gram-Schmidt algorithm \citep{GramSchmidt1994}. We then keep the top 100 basis vectors and discard the rest. A selection of these basis vectors are shown in Figure~\ref{fig:basis_vectors}, along with realisations of the binary masks generated by randomly generate coefficients.

We construct a polychromatic model of the \textsc{Toliman} optical system featuring:

\begin{itemize}
    \item A clear 125\,mm aperture diameter.
    \item A polychromatic bandpass modelled at three wavelengths: 530, 585, and 640\,nm.
    \item \ac{psf} pixel scale of 0.375\,arcseconds/pixel, $\sim 1.5\times$ Nyquist.
    \item The 4th - 10th Zernike modes (ignoring the lowest 3: piston, tip, and tilt).
\end{itemize}

\noindent We then model the expected signals from the binary star $\alpha$~Centauri~AB assuming a representative projected separation of 10\,arcseconds.%

To transform the Fisher information into a concrete optimisation objective, we construct a loss function that minimises the uncertainty of a specific scientific parameter of interest under a full forward model --- the binary separation r in this case

\begin{equation}
\text{loss}(\boldsymbol{\eta}) = \left[ \mathbf{C}(\boldsymbol{\psi}_0) \right]_{rr} = \left[ -\nabla^2_{\boldsymbol{\psi}}  \mathcal{L}(f(\boldsymbol{\eta}, \boldsymbol{\psi})) \big|_{\boldsymbol{\psi}=\boldsymbol{\psi_0}} \right]^{-1}_{rr} = \text{Var}(r \ | \ \boldsymbol{\psi}_0, \boldsymbol{\eta}),
\end{equation}

\noindent where

\begin{itemize}
    \item $\boldsymbol{\eta}$ are the design parameters, the diffractive pupil basis vector coefficients.
    \item $\boldsymbol{\psi}$ are the astrophysical and instrumental parameters over which we marginalise (e.g. binary separation $r$, flux, aberrations).
    \item $f(\boldsymbol{\eta}, \boldsymbol{\psi})$ is the differentiable forwards model simulating the system.
    \item $\mathcal{L}$ is the log-likelihood function (Poissonion, consistent with the assumption of photon noise-limited observations).
    \item $\nabla^2_{\boldsymbol{\psi}}$ is the Hessian operator applied to the $\boldsymbol{\psi}$ parameters of the model.
    \item $[\cdot]_{rr}$ denotes the component of $\boldsymbol{\psi}$ corresponding to the binary separation.
\end{itemize}

\begin{figure*}
    \centering
    \includegraphics[width=1.\textwidth]{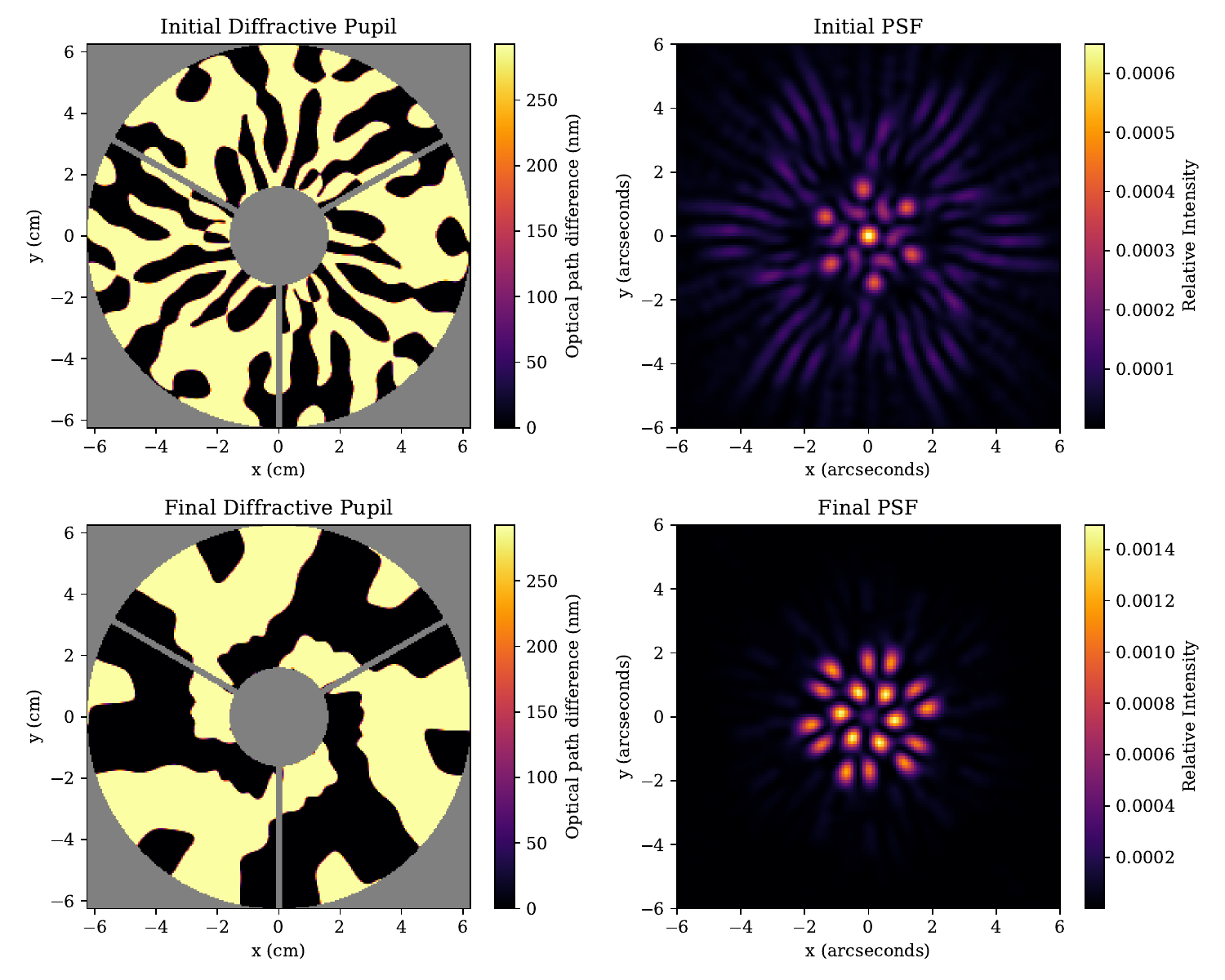}
    \caption[Diffractive Pupils and PSFs]{Diffractive Pupil - \ac{psf} pairs before and after Fisher optimisation. Top panels: Randomly initialised pupil pattern (left) consisting of binary-valued $0/\pi$ phases within the circular support of primary/secondary mirrors and 3 opaque spiders, together with its corresponding \ac{psf} (right). Bottom panels: Final pupil after optimisation (left) and its corresponding \ac{psf} (right).}
    \label{fig:masks_psfs}
\end{figure*}

We note that the likelihood is only a function of the model, as we generate the data at the input at $(\boldsymbol{\eta}, \boldsymbol{\psi})$ in order to ensure our Hessian computation produces the Fisher matrix and its inverse is therefore the \ac{crlb}. This loss function computes the covariance matrix of the parameters through the forward model, extracting the variance component corresponding to $r$, which serves as the optimisation target. Other reductions of the covariance matrix such as its trace or entropy can also be optimised if desired. Gradients of the loss with respect to the design parameters $\boldsymbol{\eta}$, ie the coefficients of the diffractive pupil basis vectors, are computed via automatic differentiation. Optimisation is performed using 50 epochs of the Adam algorithm \citep{adam}.

\begin{figure*}
    \centering
    \includegraphics[width=1.\textwidth]{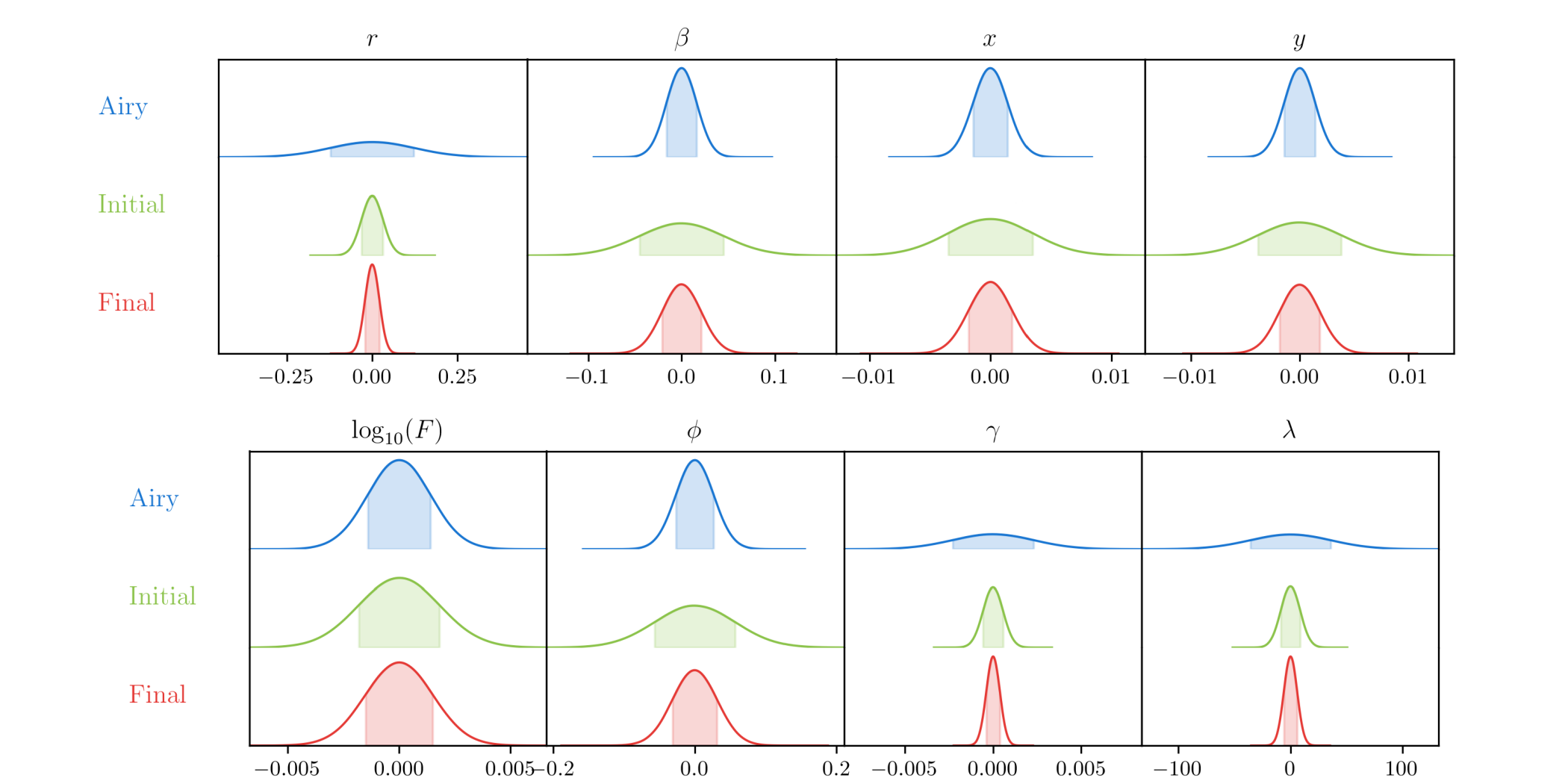}
    \caption[Posterior Distributions for Unobstructed and Diffractive Pupils]{\ac{crlb} posteriors shown as offsets from fiducial values; for flux, the offset is $\log_{10}(F/F_0)$, where $F_0$ is the fiducial flux. Optical aberrations are omitted from the plots but marginalised over. The `Airy' model has no diffractive pupil; `Initial' and `Final' denote the diffractive pupil before and after optimisation. Optimisation improves all displayed constraints relative to the initial pupil, with the greatest gains in separation $r$, pixel scale $\gamma$ and mean wavelength $\lambda$. The final model also outperforms `Airy' for these three parameters, but performs worse for the remaining displayed parameters.}
    \label{fig:post_comparison}
\end{figure*}

\begin{figure*}
    \centering
    \includegraphics[width=1.0\textwidth]{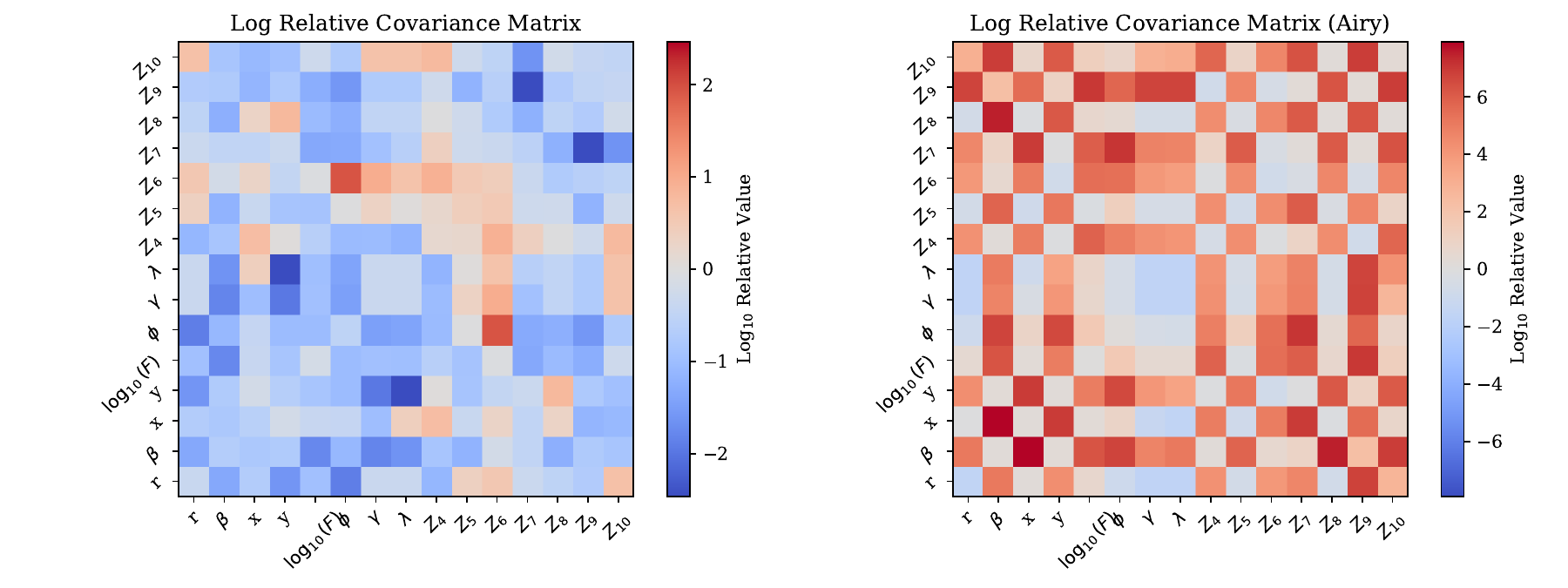}
    \caption[Relative Covariance Matrices for Pupils Before and After Optimisation]{Logarithm of absolute element-wise relative covariance matrices, i.e. $\text{log}_{10}(|C_{ij,1} / C_{ij,2}|)$. Negative values are blue and indicate a better constraint of that parameter while red flags the opposite. Left panel: The post-optimisation covariance matrix relative to the pre-optimisation covariance matrix. Right panel: The post-optimisation covariance matrix relative to the Airy-disk covariance matrix.}
    \label{fig:cov_norm}
\end{figure*}

\begin{figure*}
    \centering
    \includegraphics[width=1.0\textwidth]{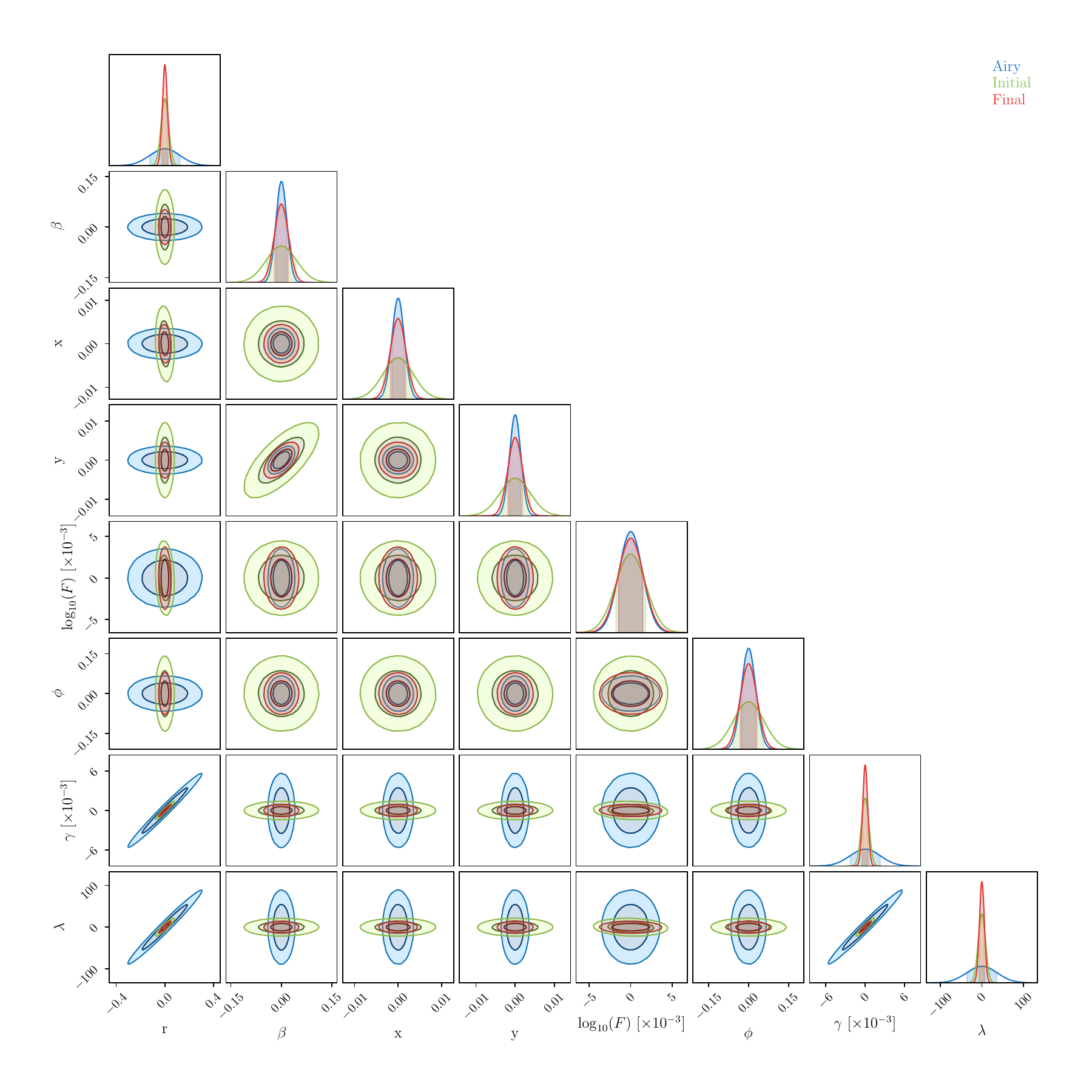}
    \caption[Reduced Corner Plot Comparing Fisher Forecasts for Pupils Pre and Post Optimization]{Reduced corner plot for the Airy-like (blue), initial (green) and optimised (red) models, showing offsets from fiducial values; for flux, the offset is $\log_{10}(F/F_0)$, where $F_0$ is the fiducial flux. All model parameters are marginalised over; optical aberrations, which behave identically across models, are omitted to keep the figure compact.}
    \label{fig:corner_plot}
\end{figure*}

The initial and final pupil-\ac{psf} pairs are shown in Figure~\ref{fig:masks_psfs}, where we can see that the final \ac{psf} is concentrated into a small number of brighter peaks near the centre, plus a series of dimmer peaks surrounding it.
Figure~\ref{fig:post_comparison} shows the parameter posteriors for both the initial and final pupils, as well as an optical system with a clear pupil producing an Airy-disk like \ac{psf} as a benchmark. We see the greatest improvement for the optimisation metric, the binary separation $r$, as well as the pixel scale $\gamma$ and mean wavelength $\lambda$. Using this Fisher optimisation approach, we improve the \ac{crlb} of the binary star separation from $\sim32$\,mas to $\sim22$\,mas, approximately a 30\% improvement from this single seed.
Figure~\ref{fig:cov_norm} compares the performance of these models to each other by visualising their relative covariance matrices ($\text{log}_{10}(|C_{ij,1} / C_{ij,2}|)$). The full corner plot is shown in Figure~\ref{fig:corner_plot}. Note that the optical aberrations have been omitted for visualisation purposes, although they were present for the optimisation and have been marginalised over.

Interestingly, the pre-optimisation model already outperforms the Airy system, implying a good choice of basis vectors for the diffractive pupil. Despite the improved performance in the binary separation metric, the diffractive pupil model is inferior in recovery of both the (x, y) position and position angle $\beta$.
Figure~\ref{fig:cov_norm} reveals some interesting properties of this optimisation process, we can see that while most elements of the covariance matrix show improved performance, some exhibit degraded performance. This is explained through the marginalisation process inherent when inverting Fisher matrices to get covariance matrices --- parameters that have little to no covariance with the parameter of interest (the binary separation $r$) can have their precision decreased in order to gain improved precision over those that are. This demonstrates an optimisation process that is fully coherent of the complex relationships between different parameters within a system.
The pupil mask design approach here offers a significant improvement in the astrometric performance of \textsc{Toliman}, and these methods will form the basis of its design and analysis, presently in progress.

\begin{figure*}
    \centering
    \includegraphics[width=0.95\textwidth]{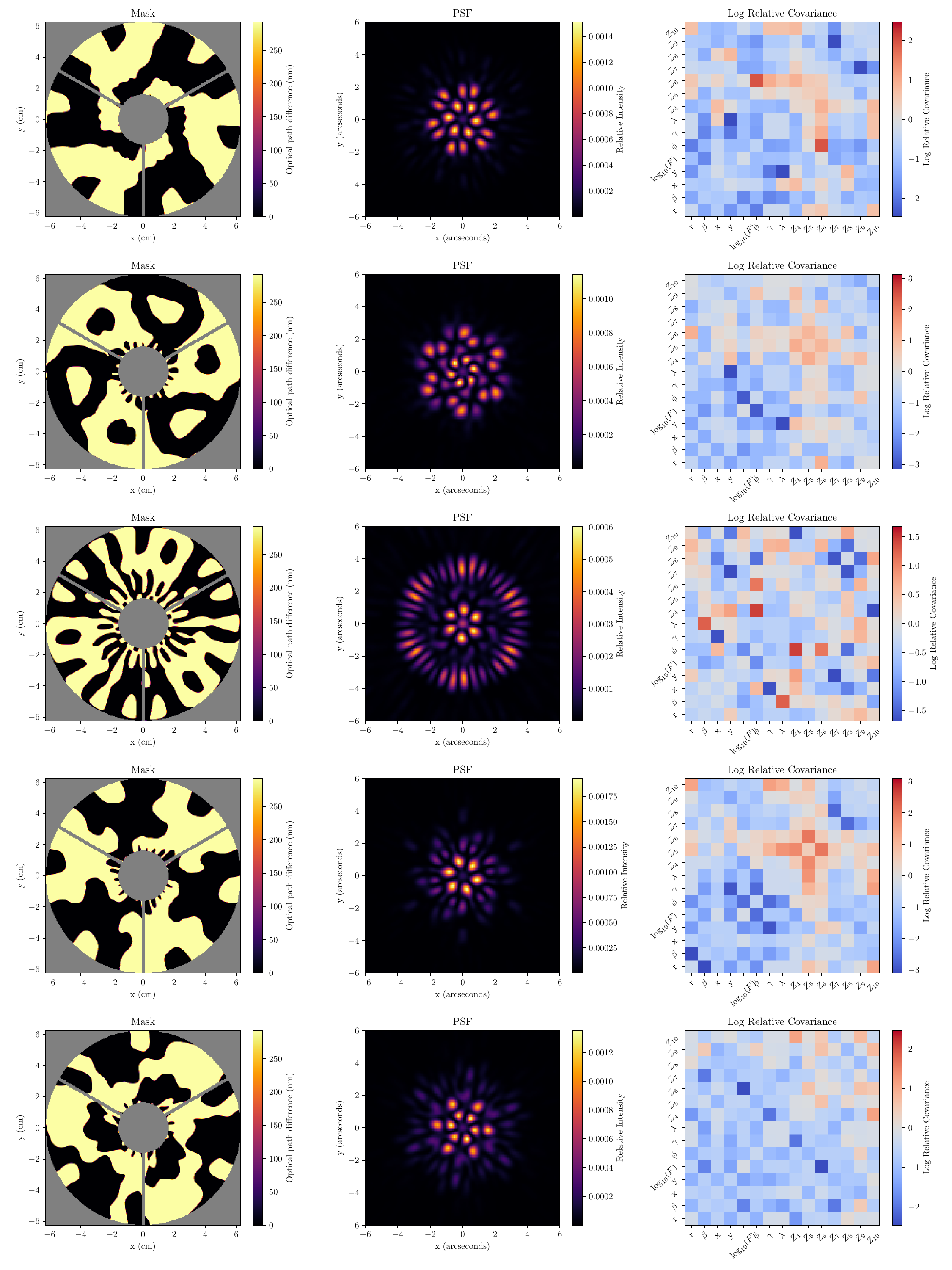}
    \caption[Optimised Diffractive Pupils and PSFs]{Summary of diffractive pupil optimisations for five different random seeds. Each row is a different seed, left panel: Final diffractive pupil, middle panel: Resulting \ac{psf}, right panel: $\log_{10}$ relative covariance matrix, normalised by the initial pre-optimisation covariance matrix.}
    \label{fig:seeds}
\end{figure*}

Given the high-dimensional nature of this problem and the complexity of the loss space introduced by the mask binarisation process, this is a non-convex problem where only local minima can be found. To ensure that the solutions found are sufficient, we repeat the optimisation process from five randomly initialised pupil patterns. Figure~\ref{fig:seeds} shows the resulting pupil designs along with their associated \ac{psf}s and relative covariance matrices, expressed as $\text{log}_{10}(|C_{ij,1} / C_{ij,2}|)$. Each of these five random seed improves the \ac{crlb} of the binary star separation over the optimisation by approximately 33\%, 48\%, 35\%, 36\% and 21\%. These results are presented as illustrative examples of the method’s capabilities, without detailed interpretation of the relative performance across different seeds.

\section{Discussion}
\label{sec:discussion}

This manuscript has detailed a methodological approach to demonstrate the practical use of Fisher-information-based experimental design methods enabled by automatic differentiation. Our aim has been to to clarify the theoretical foundations and facilitate reproducibility and extension of these methods to related problems in instrumentation and observational astronomy.
Assisting in this has been the presentation of a concrete illustrative example: furnishing the design of a complex optical component optimised for a single specific quantifiable outcome.
By first highlighting the theory and validating the appropriate conditions required, we explored how overall performance of complex systems can be analysed, and furthermore how the modelling framework can be used to optimise instrumental components. %
To maintain clarity, we chose illustrative problems that did not require the inclusion of priors. However, inclusion of simple multivariate normal priors is trivial as Fisher information is additive.

\newpage
The methods to calculate Fisher Information presented in this work provide a simple and straightforward way to explore, analyse and optimise models, however it is important to be cognisant of limitations. Firstly the Laplace approximation assumes that all posteriors are well described by multivariate normals about their peaks. This assumption will not hold in a general sense for systems with complex parameter degeneracies. While an understanding of the local topology of the posterior about the maximum likelihood estimate can be found, this can not function as a \emph{general} replacement for sampling of the posteriors via Monte Carlo methods which serve to understand the global topology of the posterior.
In many cases, the posterior distribution is either well-approximated by a multivariate normal, or a global understanding of its structure is unnecessary. In such scenarios, the second-order Laplace approximation provides sufficient insight and this is often the case for well-constrained optical instruments where local sensitivity is the dominant driver of performance. Furthermore, this work only explored the optimisation of the Fisher information about a single realisation of model parameters. In practice for an instrument like Toliman the data will span a varying range of both instrumental and astrophysical states. Accounting for this variation can be done during the optimisation process by sampling the parameters of the covariance matrix over their expected range of values, ensuring solutions robust to all states are found.

\section{Conclusion}
\label{sec:conclusions}

Progress in astronomical science depends not only on advances in hardware and observational platforms, but also on innovations in software that enhance instrument design and data analysis. Exoplanetary science, in particular, places some of the most stringent demands on both instrumentation and inference due to the intrinsically faint, noise-dominated signals that require careful calibration and characterisation. Autodiff has already demonstrated profound utility in improving astrophysical measurements and calibration techniques \citep{Liaudat2023}. Modern frameworks such as \jax enable a new class of scientific software that builds on autodiff to offer powerful extensions to existing tools. This manuscript aims to showcase the modelling strategies made possible by adopting differentiable software paradigms such as \dlux that treat autodiff not as an afterthought, but as a foundational design principle. By replacing hand-derived analytic equations with programmatic differentiation, these tools allow complex models that incorporate realistic optics and detector effects to be optimised and explored using the same physical principles, but with greater flexibility and computational tractability.

In this manuscript we have explored new avenues for analysis and design of optical systems harnessing the stable calculation of Fisher matrices empowered by autodiff. This method yields results in agreement with analytically derived expressions when identical assumptions are enforced, and furthermore it can be used to explore extensions to a more relaxed set of assumptions. The framework provides new ways to probe the effects of different instrumental architectures on the recovery of science outcomes, and straightforward algorithms to optimise experimental design targeting constraint of specific astrophysical parameters without detailed pen-and-paper analysis.

Differentiable methods also enable inference to be performed on complex posteriors. In cases where distributions are non-Gaussian due to complex parameter degeneracies, approximate inference methods become essential. Stochastic variational inference \citep{hoffman2013stochastic} enables parameter degeneracies resulting in non-Gaussian posteriors to be approximated by minimising the divergence between the inferred and true posterior. Extending this, automatic differentiation variation inference \citep{kucukelbir2016automatic} utilises the derivatives of a model to explore the posterior parameter space efficiently, and can be applied directly to differentiable optical models in order to address unavoidable parameter degeneracies present in complex data sets.

These methods offer a new framework not only for analysing observational data, but also for informing the design of future instruments employing statistically grounded optimisation criteria described by the Fisher information.
As one particularly promising use case, future coronagraphic instrument design based on autodiff may be made more robust to realistic noise processes such as low-order wavefront error \citep{Currie2018} or the low wind effect \citep{Milli2018}. %
With these tools, novel architectures can be rapidly explored and optimised with respect to evidence ratios, robustness to optical aberrations, or spectral parameter estimation. The same principles can be applied to other instruments such as spectrographs for radial velocity planet detection, enabling joint end-to-end optimisation of optical systems and observational regimes.

The methods presented were implemented using \dlux, however they are not inherently tied to any specific software package. The central contribution of this work lies in the application of automatic differentiation for optical system design and uncertainty forecasting. We anticipate that similar approaches can be adapted within other autodiff-capable frameworks, and future work may benefit from comparative benchmarking across such platforms.

\section{Code, Data, and Materials}
\label{sec:code}

As part of our commitment to open science, we have released \dlux as an open source package under a BSD three-clause at~\href{https://github.com/LouisDesdoigts/dLux}{github.com/LouisDesdoigts/dLux}.
Furthermore an accompanying Jupyter notebook that produces all results and figures in this paper is publicly hosted at~\href{https://github.com/LouisDesdoigts/FIM_tutorial/blob/main/tutorial.ipynb}{github.com/LouisDesdoigts/FIM\_tutorial}.
The accompanying code repository allows for replication of results and adaptation of the methods to a variety of optical modelling tasks.

\section*{Acknowledgements}

Financial and logistical support for this research program were provided by the Breakthrough Prize Foundation as a part of the Breakthrough Watch initiative.

We acknowledge and pay respect to the traditional owners of the land on which the University of Queensland, Macquarie University, and the University of Sydney are situated, upon whose unceded, sovereign, ancestral lands we work. We pay respects to their Ancestors and descendants, who continue cultural and spiritual connections to Country.

This research made use of \textsc{NumPy} \citep{numpy}; Matplotlib \citep{matplotlib}; \textsc{Jax} \citep{jax}; \numpyro \citep{Phan2019}; \texttt{equinox} \citep{kidger2021equinox}; \optax \citep{optax2020github}; and ChainConsumer \citep{Hinton2016}.

\end{spacing}

\bibliographystyle{apsrev4-1}

\bibliography{output}

\end{document}